# Robustness of binary stochastic neurons implemented with low barrier nanomagnets made of dilute magnetic semiconductors

Rahnuma Rahman and Supriyo Bandyopadhyay
Department of Electrical and Computer Engineering
Virginia Commonwealth University, Richmond, VA 23284, USA

**ABSTRACT**

Binary stochastic neurons (BSNs) are excellent hardware accelerators for machine learning. A popular platform for implementing them are low- or zero-energy barrier nanomagnets possessing in-plane magnetic anisotropy (e.g. circular disks or quasi-elliptical disks with very small eccentricity). Unfortunately, small geometric variations in the lateral shapes of such nanomagnets can produce large changes in the BSN response times if the nanomagnets are made of common metallic ferromagnets (Co, Ni, Fe) with large saturation magnetization. Additionally, the response times are also very sensitive to initial conditions. Here, we show that if the nanomagnets are made of dilute magnetic semiconductors with much smaller saturation magnetization, then the variability in their response times (due to shape variations and variation in the initial condition) is drastically suppressed. This significantly reduces the device-to-device variation, which is a serious challenge for large scale neuromorphic systems.

## I. Introduction

ARTIFICIAL intelligence (AI) platforms often employ spiking neural networks (SNN) activated by binary stochastic impulses to execute machine learning tasks [1, 2]. The common building block of SNNs is a binary stochastic neuron (BSN) that has two distinct output states; the BSN will output either state with a probability determined by a specific function of an input impulse. BSN states can also represent Ising spins and are useful hardware platforms for Ising machines that can solve combinatorial optimization problems.

Recently, it has become popular to implement BSNs with low (or zero) energy barrier nanomagnets (LBMs) possessing in-plane magnetic anisotropy because of their superior energy efficiency and relatively fast response times [3-5]. The magnetization vectors of these LBMs, which are circular or nearly circular disks, fluctuate randomly at room temperature owing to thermal perturbations and the fluctuating magnetization (with suitable hardware design) can produce a binary state $m_i$ (-1 or +1) at time step $(n+1)$ given by

$$m_i(n+1) = \mathrm{sgn}\left[\tanh\left(I_i(n)\right) - r_i\right], \tag{1}$$

where $I_i$ is a dimensionless input spin current that biases the output and $r_i$ is a random number uniformly distributed between -1 and +1. In the absence of $I_i$, the outputs $-1$ and $+1$ are equally likely. A positive $I_i(n)$ makes $+1$ more likely, and a negative $I_i(n)$ makes $-1$ more likely. Each BSN described by Equation (1) receives its input from a weighted sum of other BSNs obtained from a synapse $I_i(n) = \sum_j W_{ij} m_j(n)$. A wide variety of problems can be solved by properly designing or learning the weights $W_{ij}$, e.g. classification problems [6], constrain satisfaction problems [7], generation of cursive letters [8], etc.

A critical parameter for a BSN implemented with LBMs is the "correlation time" $\tau_c$ which is the full-width-at-half maximum (FWHM) of the temporal decay characteristic of the auto-correlation function of the magnetization fluctuations [4]. This quantity determines the BSN's *response speed*. LBMs with in-plane anisotropy typically have a value of $\tau_c$ that is about two orders of magnitude smaller than LBMs with perpendicular magnetic anisotropy [4] and are therefore favored because they lend themselves to *faster* circuit operation. The correlation time, however, is very sensitive to small geometric variations if LBMs with in-plane magnetic anisotropy are realized with magnetic materials possessing *large saturation magnetization*, e.g. Co, Ni or Fe [9, 10]. The reason for this is that the shape anisotropy barrier $E_b$ in a nanomagnetic disk that is not a perfect circle but is slightly elliptical, as shown in Fig. 1, is given (within the macrospin approximation) as

$$E_b = \frac{\mu_0}{2} M_s^2 \Omega \left( N_{d-xx} - N_{d-yy} \right), \qquad (2)$$

where $\mu_0$ is the permeability of free space, $\Omega$ is the nanomagnet volume, $N_{d-xx}$ and $N_{d-yy}$ are the demagnetization factors along the minor and major axes of the ellipse, respectively, and $M_s$ is the saturation magnetization. The expressions for the demagnetization factors are [11]

$$N_{d-xx} = \frac{\pi}{4}\left(\frac{t}{a}\right)\left[1 + \frac{5}{4}\left(\frac{a-b}{a}\right) + \frac{21}{16}\left(\frac{a-b}{a}\right)^2\right]$$

$$N_{d-yy} = \frac{\pi}{4}\left(\frac{t}{a}\right)\left[1 - \frac{1}{4}\left(\frac{a-b}{a}\right) - \frac{3}{16}\left(\frac{a-b}{a}\right)^2\right] \qquad (3)$$

where $a$ = major axis dimension, $b$ = minor axis dimension and $t$ = thickness of the elliptical nanomagnet. For a perfect circle, $a = b$ and hence $E_b = 0$. For an ellipse, $a > b$ and hence $E_b > 0$.

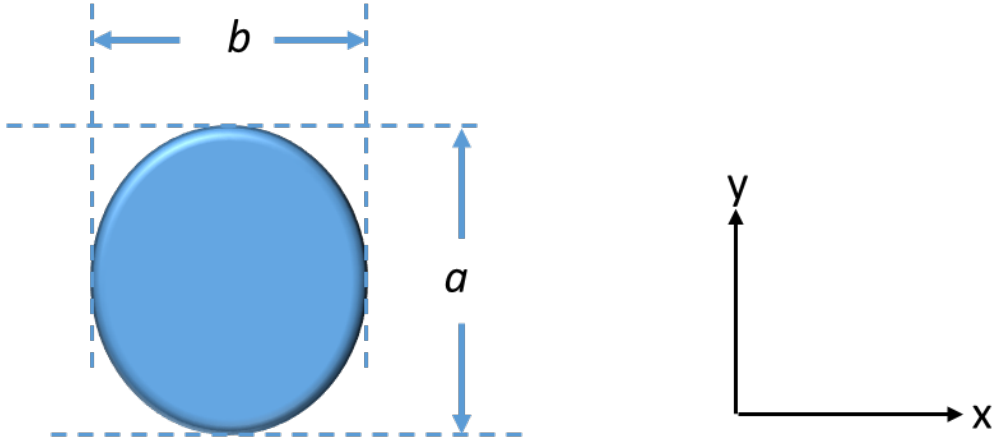

Fig. 1: A nanomagnet shaped like a slightly elliptical disk.

It is clear from Equations (2) and (3) that a slight change in the nanomagnet dimension $a$ or $b$ can cause a large change in the energy barrier $E_b$ if the saturation magnetization of the nanomagnet $M_s$ is large. A material like Co has a saturation magnetization of $10^6$ A/m, whereas a dilute magnetic semiconductor like GaMnAs has a saturation magnetization of only $5 \times 10^3$ A/m [12] at room temperature, which is more than two orders of magnitude smaller. Hence, everything else being the same, a GaMnAs nanomagnet will have an energy barrier that is roughly four orders of magnitude smaller than a Co nanomagnet. Hence, the same change in a nanomagnet dimension (*a* or *b*) will cause a much smaller change in the energy barrier if the nanomagnet is made of GaMnAs than if it is made of Co. Thus, a BSN implemented with a GaMnAs LBM is expected to be much more robust against small geometric variations than one implemented with a Co nanomagnet.

## II. Results

We simulate the magneto-dynamics of two sets of nanomagnets made of Co and GaMnAs, respectively, using the stochastic Landau-Lifshitz-Gilbert equations and different initial conditions (i.e. different initial magnetization orientations) [9]. This yields the magnetization vector as a function of time and therefore the autocorrelation plots of any component of the magnetization as a function of the delay [9], from which we can extract the correlation time $\tau_c$ for that component. Each of the two sets of nanomagnets consists of a slightly elliptical one of major axis 100 and minor axis 99 nm, and a perfectly circular one of diameter 100 nm. The magnet thickness is 6 nm. The in-plane shape anisotropy energy barrier in the case of the slightly elliptical Co nanomagnets is ~4 kT, while in the case of GaMnAs nanomagnets, it is $10^{-4}$ kT at room temperature. The Gilbert damping factor in GaMnAs is 0.01 [13], similar to that of Co.

We simulate the magneto-dynamics in the presence of thermal noise at room temperature (300 K), taking into account *spin inertia* which causes nutational dynamics [14]. The nutational dynamics lasts for a duration $\tau$ (which depends on the magnetic vector's moment of inertia and damping) and we consider three values of $\tau = 1$, 10 and 100 ps. The detailed procedures for calculation can be found in ref. [9] and not repeated here to avoid redundancy.

We calculate the autocorrelation functions of the magnetization components along the minor and major axes of the nanomagnets, $C_x(t') = \int_0^\infty m_x(t) m_x(t+t') dt$ and $C_y(t') = \int_0^\infty m_y(t) m_y(t+t') dt$ with two sets of initial conditions: $m_x(0) = 0.995; m_y(0) = 0.095; m_z(0) = 0.031$ (initial orientation along the minor axis) and $m_x(0) = 0.095; m_y(0) = 0.995; m_z(0) = 0.031$ (initial orientation along major axis). We then plot the autocorrelation functions as a function of the delay *t'* for the Co and GaMnAs nanomagnets for both slightly elliptical and circular nanomagnets and for both initial conditions in Fig. 2. These characteristics are averaged over 1000 simulation runs which is equivalent to ensemble averaging over 1000 identical nanomagnets for each plot (the averaging is necessary since thermal fluctuations are random). We have ascertained that a larger ensemble size does not change any result perceptibly.

We see no significant dependence on spin inertia since the plots do not change appreciably for the three different values of $\tau$, but this is expected since the nutational duration $\tau$ is much smaller than the correlation time $\tau_c$ in all cases. There are situations when the nutational dynamics can affect outcomes which emerge long after nutation has ceased [15], but this is not such a case. Clearly, the fluctuation magneto-dynamics does not retain any memory of the nutation long after it ceases to exist.

The interesting observation is that in the case of GaMnAs, unlike in the case of Co, the autocorrelation decay plots are *not very sensitive* to small deviations in shape (perfectly circular versus slightly elliptical), nor are they particularly sensitive to the initial orientation of the magnetization for the case of the slightly elliptical nanomagnet. The latter feature can be understood easily. If the initial orientation of the magnetization is along the major axis of the ellipse, which is the easy axis, the magnetization is in a stable state and it will take a long time for thermal perturbations to destabilize it *if the energy barrier is large.* On the other hand, if the magnetization is initially along the minor axis (hard axis), it is at the maximally unstable location and thermal perturbations will very quickly destabilize it if the *energy barrier is large*. That is why in the case of Co (large energy barrier owing to large saturation magnetization) $\tau_c$ is ~20 ns if the initial orientation is along the major axis and only ~1.5 ns if it is along the minor axis. The difference is more than an order of magnitude. This large difference (and hence variability) would mandate an initialization step where the initial magnetizations of all BSNs are forcibly aligned in the same direction with a magnetic field, which is an inconvenience. Second, we notice that even a 1% change in a lateral dimension can change the correlation time $\tau_c$ by a factor of 4! This would demand unrealistic fabrication tolerance. On the other hand, in the case of the GaMnAs nanomagnets, $\tau_c$ varies between 5.95 and 6.23 ns for all cases, showing that the variability is dramatically suppressed. Neither is the correlation time sensitive to small geometric variations, nor is it sensitive to initial conditions. This is, of course, a consequence of the much smaller energy barrier in the case of GaMnAs compared to Co.

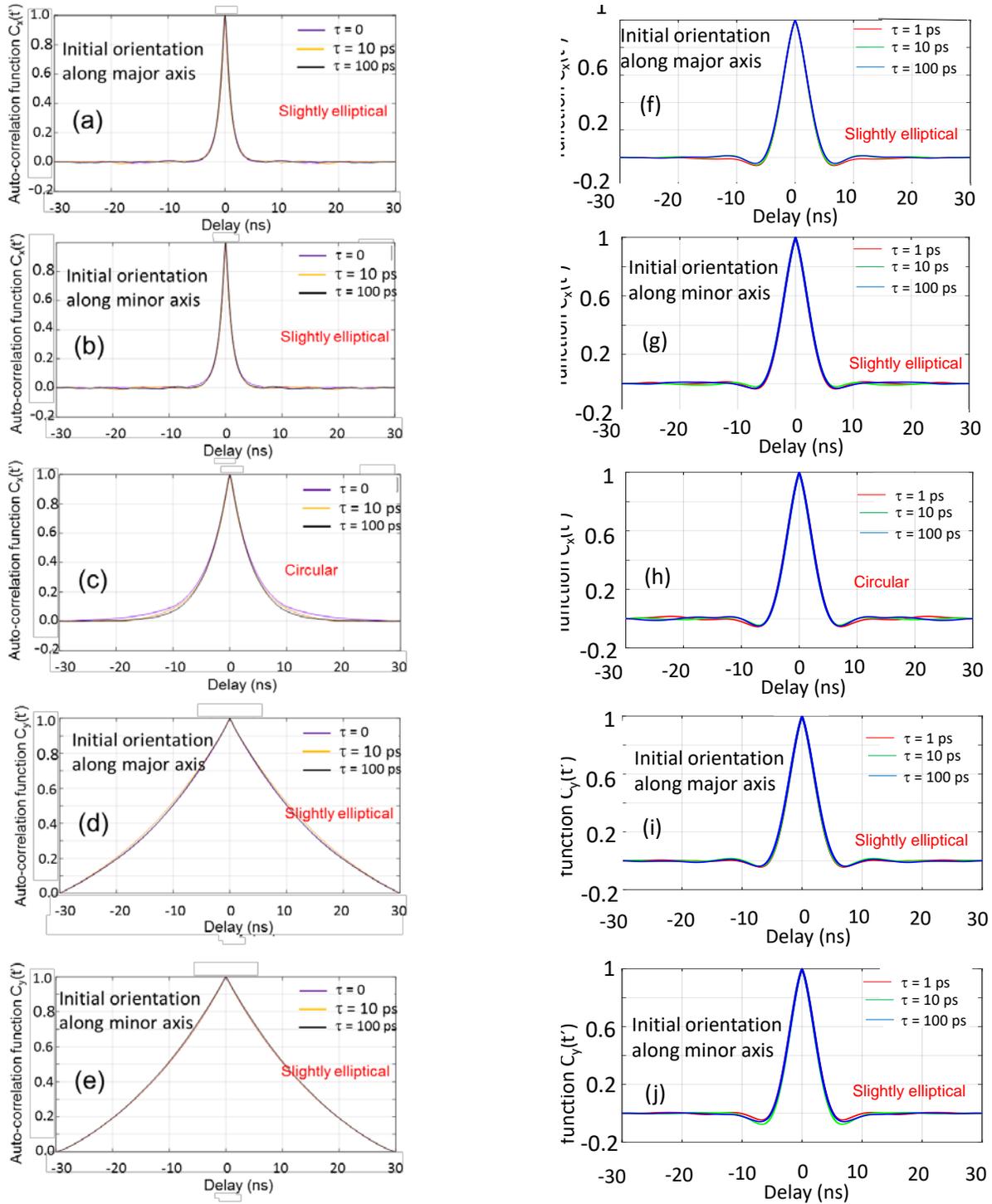

Fig. 2: Left column (a) – (e) are for Co nanomagnets of slightly elliptical and circular shapes for different initial orientations of the magnetizations and right column (f) – (j) are the corresponding plots for GaMnAs nanomagnets. Figures (a) – (e) are reproduced from [9].

### III.     Conclusion

Device to device variability in BSNs, and neuromorphic circuits in general, is a serious problem and countermeasures have been proposed to alleviate it. One recent suggestion was to use hardware-aware in-situ learning [16]. Here, we propose a simpler solution; replacement of a magnetic material with a large saturation magnetization with one with smaller saturation magnetization, such as a dilute magnetic semiconductor like GaMnAs. This reduces the sensitivity of the correlation time (response time) of BSNs to small geometric variations and also initial conditions.

**Acknowledgement:** This work was supported in part by the National Science Foundation under grants CCF-2006843 and CCF-2001255.